\documentclass[aps,twocolumn,amssymb, showpacs]{revtex4}
\usepackage{epsfig}

\begin{document}

\title{An energy landscape model for glass-forming liquids in
three dimensions}
\author{Ulf R. Pedersen, Tina Hecksher, 
        Jeppe C. Dyre, and Thomas B. Schr{\o}der}
\email[]{E-mail: tbs@ruc.dk}

\affiliation{
Department of Mathematics and Physics (IMFUFA),\\
DNRF centre "Glass and Time",\\
Roskilde University, Postbox 260, DK-4000 Roskilde, Denmark}
\date{\today}

\begin{abstract}
We present a three-dimensional lattice-gas model with trivial 
thermodynamics, but nontrivial dynamics. The model is characterized 
by each particle having its own
random energy landscape. The equilibrium dynamics of the model 
were investigated by continuous time Monte Carlo
simulations at two different densities at several temperatures. At high 
densities and low temperatures the model captures the important 
characteristics of viscous liquid dynamics. We thus observe non-exponential 
relaxation in the self part of the density auto-correlation
function, and fragility plots of the self-diffusion constant and 
relaxation times show non-Arrhenius behavior. 
\end{abstract}

\maketitle

\section{Introduction}
A mechanical system of $N$ spherically symmetric particles is completely characterized by
its so-called energy landscape, the graph of the potential energy
function $U({\bf r}_1,...,{\bf r}_N)$ in $3N+1$ dimensions. As
suggested by Goldstein in his pioneering 1969 paper \cite{gol69}, the
energy landscape is particularly useful for elucidating the dynamics of
highly viscous liquids. This is because viscous liquid dynamics are
dominated by jumps over barriers much larger than $k_BT$;  most time is
spent on vibrations around local energy minima of the landscape.
However, it was only after the  work of Stillinger and Weber
in the 1980's \cite{stillinger83,stillinger88} and the enormous growth
in use of computer
simulations in the 1990's that the energy landscape became a dominant
paradigm in the study of viscous liquids \cite{
speedy,dasgupta,sciortino,heuer,sastry,Schulz,ruocco,schroder}.
For recent reviews see, e.g., \cite{wal03,Sciortino05}.

It is difficult to imagine a complex high-dimensional landscape, but an
obvious idea is to assume that there is an element of randomness in the
landscape. In this philosophy one follows Wolynes, who argued that
some phenomena occurring in a specific complex system are typical of
those that occur in most systems chosen randomly out of an ensemble of
possible systems \cite{wol92}.

A possible disordered landscape consists of a high-dimensional lattice
with random, uncorrelated energies chosen, e.g.,
according to a Gaussian, with nearest-neighbor Metropolis dynamics. 
This model, which has trivial thermodynamics, has
been shown to reproduce a number of observed properties of viscous
liquids, and the low-temperature dynamics of the model are understood
to be dominated by site percolation \cite{bas87,dyr95}. 
However, the model does not have a meaningful thermodynamic limit;
if the distribution of energies is chosen such that the mean energy
is extensive, the relaxation times are not intensive.
The problem, of course, is that a single nearest neighbor jump
on the lattice changes the energy by an extensive amount, effectively
corresponding to a complete rearrangement of all molecules.
Another problem is how dimensionality is reflected in the energy landscape.
Many condensed matter systems behave differently in two and three
dimensions. If this applies also for viscous liquids, it must
somehow be reflected in the landscape.

 The question
we consider here is: Is it possible to construct a sensible 'generic'
random landscape model? Such a model should obey the following
requirements:

\begin{enumerate}
  \item It should have a well-defined  thermodynamic limit, 
        i.e., extensive average energy and intensive relaxation times.
  \item It should reflect the dimensionality of space.
  \item All sites should be statistically equivalent, thereby ensuring 
        translational invariance on the average. 
\end{enumerate}

\section{The model}
The energy landscape is attractive because it abstracts from three
dimensions. Nevertheless, we would like to suggest that the simplest
way to have a  model obeying the requirements listed above 
is to return to three dimensions:

Consider a lattice gas in three dimensions. If random 
energies are assigned to the lattice sites, the system is described by Fermi
statistics. This corresponds to particles in an external random
potential, thus with
no translational invariance and only the trivial self exclusion 
particle-particle interaction. A simple modification turns this model into a highly
nontrivial model, namely to assume that each particle has its own
energy landscape, i.e., the energy of the system is given by (where ${\mathbf
r}_i$ is the lattice position of particle $i$, and $\delta$ is the Dirac 
$\delta$-function): 
\begin{eqnarray}
  \label{eq:Hamiltonian}
  E  = \sum_{i} \epsilon_i({\mathbf r}_i) 
    +  \sum_{i\neq j} 
       \delta({\mathbf r}_i-{\mathbf r}_j) 
\end{eqnarray}

The first term is the energetic interaction; for each
lattice site, ${\mathbf r}$, and each particle, $i$, the energy, 
$\epsilon_i({\mathbf r})$, is chosen randomly from a probability 
distribution, $p(\epsilon)$. $\epsilon_i({\mathbf r})$ and  
$\epsilon_j({\mathbf r})$ are generally different random numbers 
if $i\neq j$, i.e., particles have different energy landscapes.
The second term in Eq. (\ref{eq:Hamiltonian}) is the self 
exclusion particle-particle interaction; no more than one 
particle is allowed at each lattice site.

Now all lattice sites are \emph{statistically} equivalent. 
The model allows calculation of pressure and chemical potential, 
and has extensive 
thermodynamics and intensive relaxation times. In particular we 
expect that at high densities, $\rho \equiv N/V$, there  will be a 
jamming effect slowing down the dynamics considerably.

When comparing the simulation results of this model to results from 
molecular dynamics (MD) simulations, one should keep in mind that the model 
does not include the high frequency vibrations associated with
"cage-rattling". The dynamics that are modeled here are the so called "inherent
dynamics" \cite{schroder}, i.e., the result of mapping the true dynamics 
onto a series of inherent structures (local minima in the $3N-1$
dimensional energy landscape). 

We make the 
simplest possible choice for the probability distribution $p(\epsilon)$:
the Box distribution [$p(\epsilon) = 1, 0\leq\epsilon<1$]. In this
case the mean system energy per particle is easily found to be ($\beta
\equiv 1/k_BT$):
\begin{eqnarray}
  \label{eq:Energy}
  \frac{\left< E \right> }{N} = 
    \frac{1}{\beta} - \frac{1}{\exp(\beta)-1}
\end{eqnarray}
At low temperatures, we thus get $\left< E \right> = Nk_BT$.

\section{Simulation details}

The model was simulated on a three dimensional $L\times L\times L$ 
cubic lattice using the N-fold way kinetic Monte Carlo Method 
\cite{Nfold1,Nfold2} with continuous time. We use Metropolis transition 
rates with local Monte Carlo moves; if a particle jumping to a 
nearest-neighbor site brings the system from state $i$ to state $j$, the
associated  transition rate is given by:
\begin{eqnarray}
  \label{eq:Metropolis}
  \Gamma(i\rightarrow j) 
    = \min[\Gamma_0,\Gamma_0\exp(-\beta(E_j - E_i))]
\end{eqnarray}
Our length and time units are defined by setting the lattice unit
$a\equiv 1$, and the fastest transition rate $\Gamma_0\equiv 1$.

Since each particle has its own energy landscape, 
the number of different site energies are given by 
$N\times L^3 = \rho\times L^6$. Storing  
these numbers in memory would put a severe constraint 
on how large systems we could simulate. Instead we utilize the
'ran4' random number generator \cite{Press92} in the following 
way: each particle is assigned a 'particle-seed', and 
when needed this is used together with the appropriate  
site-index as input to 'ran4', which performs a series of bit 
operations to produce a uniform random deviate 
in the range $0.0$ to $1.0$.

The simulations are carried out with two sets of periodic boundary 
conditions, one for each term in eq. \ref{eq:Hamiltonian}. We denote
by $L$ the  lattice length associated with the  particle-particle interactions.
The usage of 'ran4' makes it possible to use much larger energy landscapes: 
the lattice site index used as input to 'ran4' is a 32-bit integer, 
and for the energy landscapes we can therefore use a side-length
of approximately $1000$ (we use $L$ times an integer), i.e. for all practical 
purposes (at least in this paper) each particle has an energy 
landscape that is infinite.

As mentioned above, we impose locality on the Monte Carlo
moves; particles can only jump to (vacant) nearest-neighbor sites. 
Relaxing this
requirement (letting a Monte Carlo move consist of a random particle
interchange with a random particle/hole) gives us an efficient way 
to equilibrate the model: For all the state points investigated here 
the  characteristic time for equilibration was found to be less than
$5$ time units. Equilibration runs were done for a time period
of $1000$ time units.

\section{Simulation Results}

\begin{table}
\center{
\begin{tabular}{|c|c|c|c|} \hline
 $\rho$ & $\rho_h \equiv 1-\rho$ & $\beta$-values  & $L$ \\ \hline
 0.992  & $8\times10^{-3}$       & $0,2,4,...,16$  & 10 \\ \hline
 0.999  & $1\times10^{-3}$       & $0,1,2,...,13$  & 20 \\ \hline
 \end{tabular}
}
\caption{Parameters used in simulations. $\rho \equiv N/L^3$, $\rho_h$
  is the density of holes (unoccupied sites). $\beta \equiv 1/k_BT$
}
\label{tab:StatePoints}
\end{table}

Two densities were simulated, each at a range of $\beta$-values, 
see table \ref{tab:StatePoints}. Our simulations
agree with the analytical expression for the mean energy 
(equation \ref{eq:Energy}). Reported results are averages over 
8 independent simulations (8 different energy landscapes), and error-bars are
estimated from fluctuations between these 8  simulations.

\begin{figure}
\resizebox{8.5cm}{!}{\includegraphics{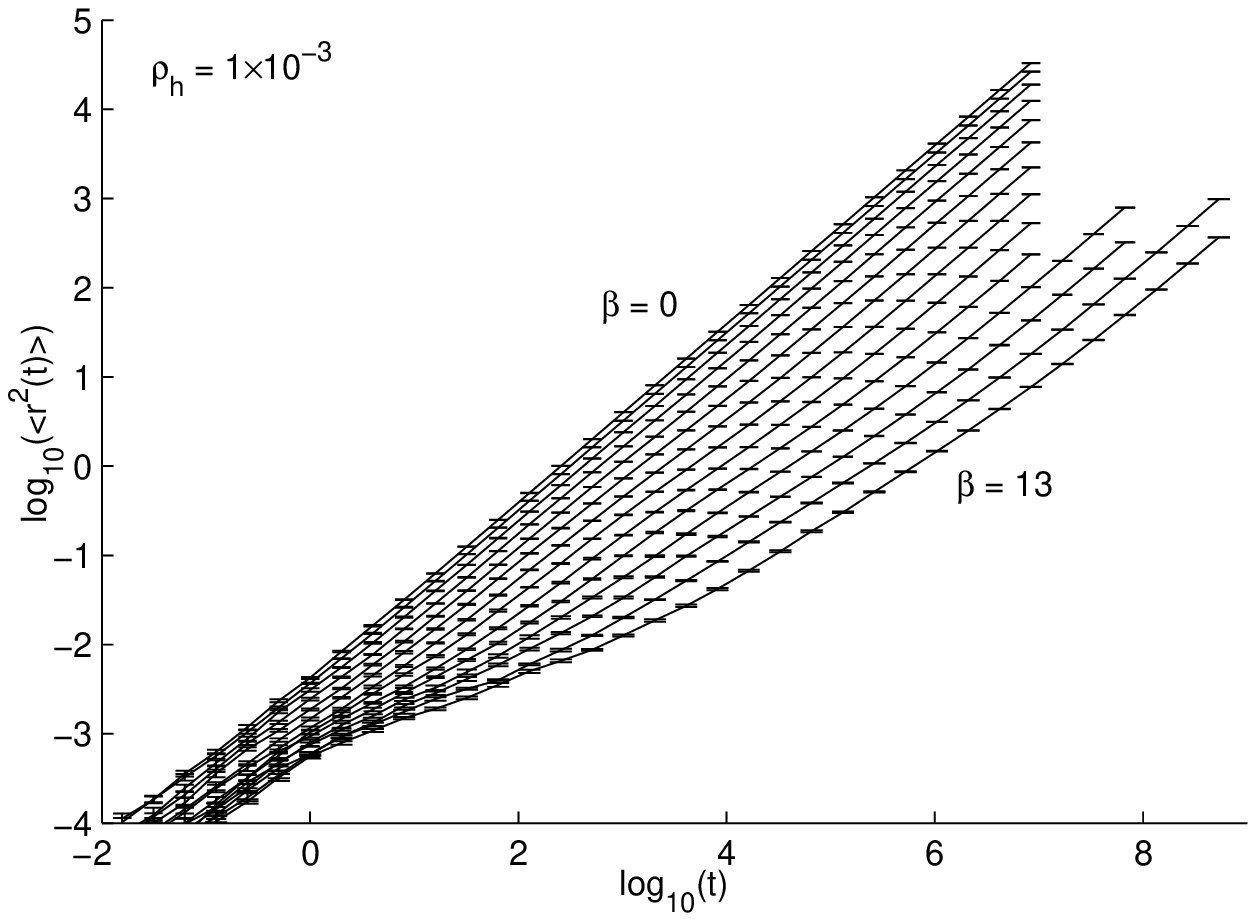}}
\resizebox{8.5cm}{!}{\includegraphics{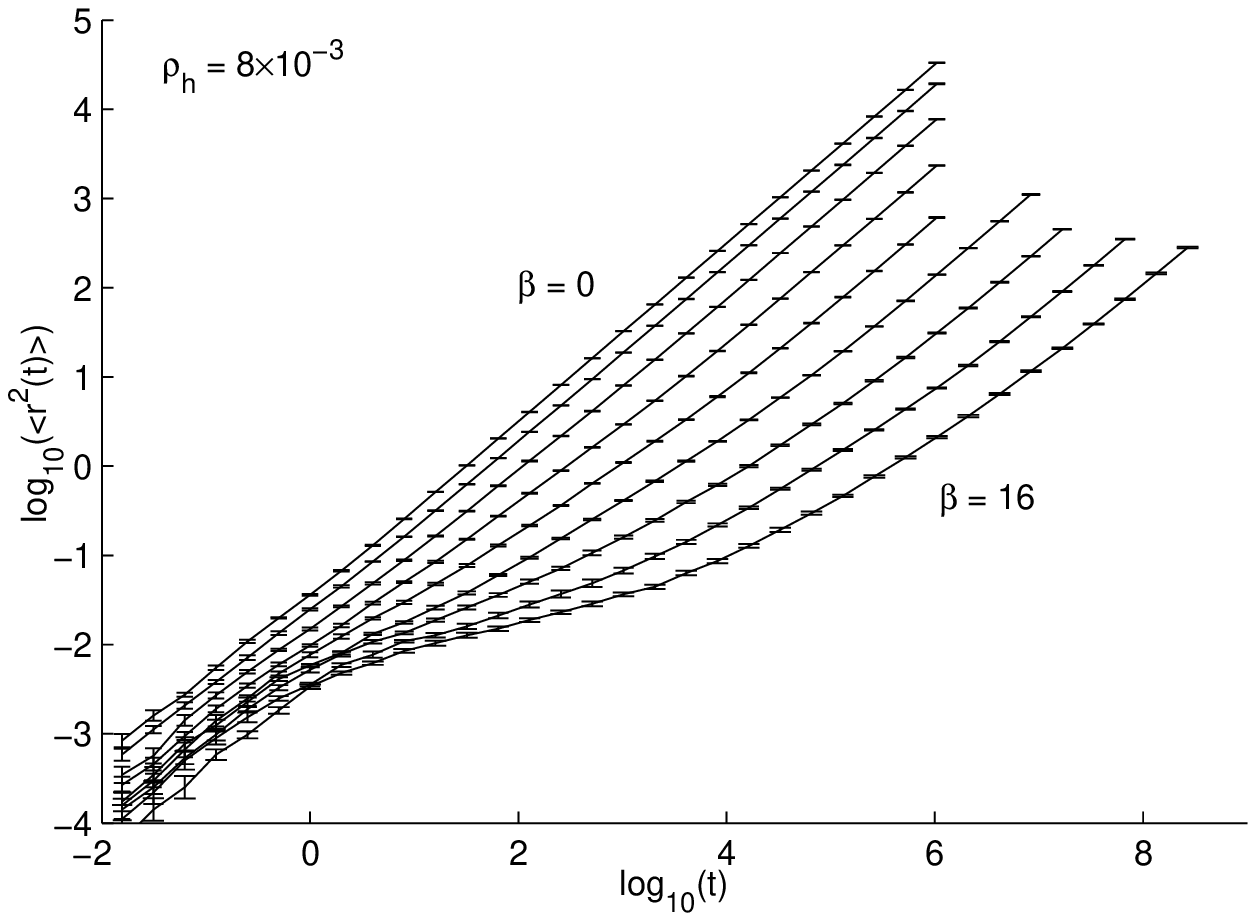}}
\caption{Mean-square displacement, $\left<r^2(t)\right>$, for 
$\rho_h = 1\times 10^{-3}$ (upper panel) and $\rho_h = 8\times 10^{-3}$
(lower panel). See table \ref{tab:StatePoints} for further details. 
Data points are connected by straight lines. Error-bars indicate $95\%$ confidence interval in $\left<r^2(t)\right>$ estimated from fluctuations between
8 independent samples (uncertainties only discernible at short times).
}
\label{fig:R2}
\end{figure}

\subsection{Mean-square displacement}
Fig. \ref{fig:R2} shows the mean-square displacement, $\left<r^2(t)\right>$, 
in a log-log plot. These results look similar to what is found in 
MD simulations of viscous liquids (see eg. \cite{Kob95}): At long
times the dynamics is diffusive ($\left<r^2(t)\right> \propto t$), and
this diffusive regime is preceded by a plateau that develops 
as the system is 
cooled. In MD simulations this developing plateau is attributed to
"cage rattling": particles vibrating in a cage consisting of the
nearest neighbors. In this regime particles move
considerably less than the inter-particle distance. This "MD scenario"
is obviously \emph{not} what happens in this model;  as discussed above the vibrations 
on length scales shorter than the inter-particle distance are not
included in the model. Here the
developing of a plateau means that after a particle has jumped
to a nearest-neighbor site, the probability for jumping  
back to where it came from is (on average) larger than the probability
for jumping to a new lattice site. This  leads to a slowing down
of the dynamics compared to diffusion dynamics. Only  when
particles have jumped several times is this correlation between jumps
lost whereupon the dynamics become diffusive. 
At short times ($t<1$) a regime with 
$\left<r^2(t)\right> \propto t$ is seen - this is simply a consequence of 
the time scale being so short that particles never jump more than
once.


\begin{figure}
\resizebox{8.5cm}{!}{\includegraphics{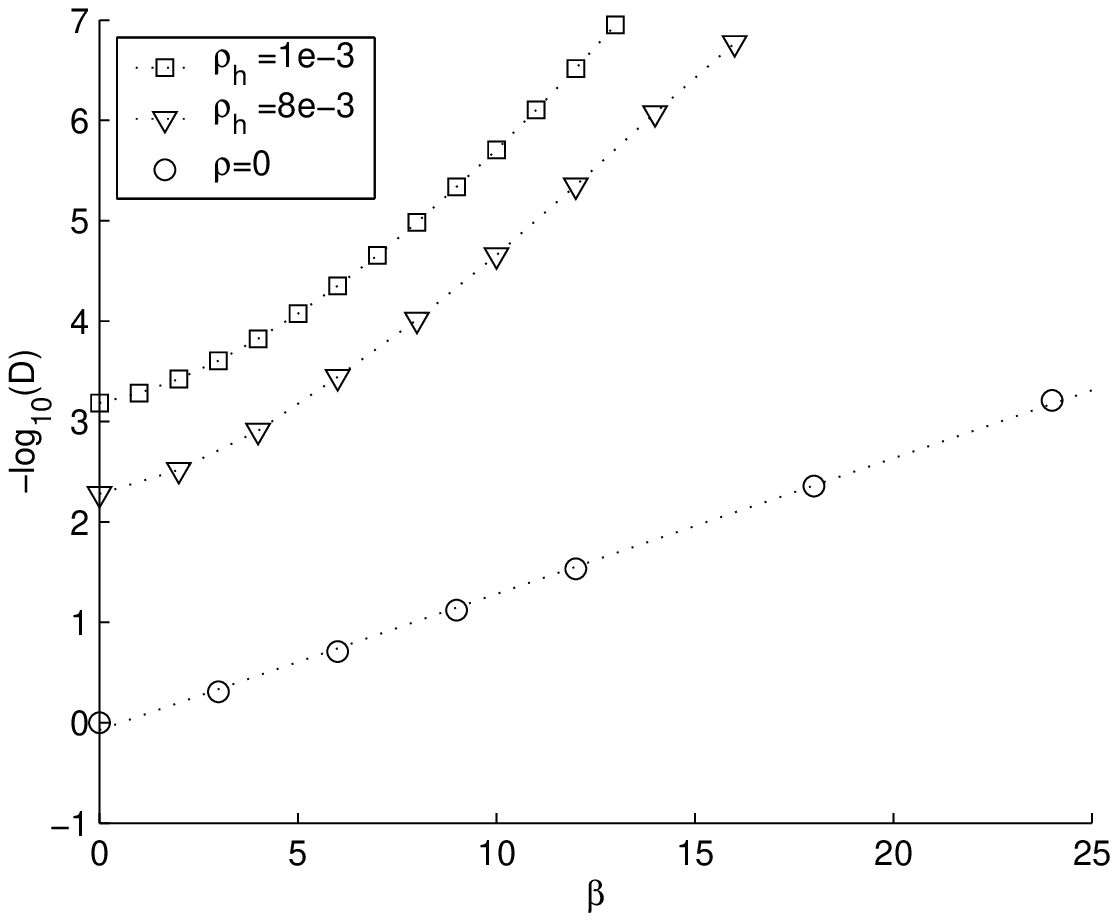}}
\resizebox{8.5cm}{!}{\includegraphics{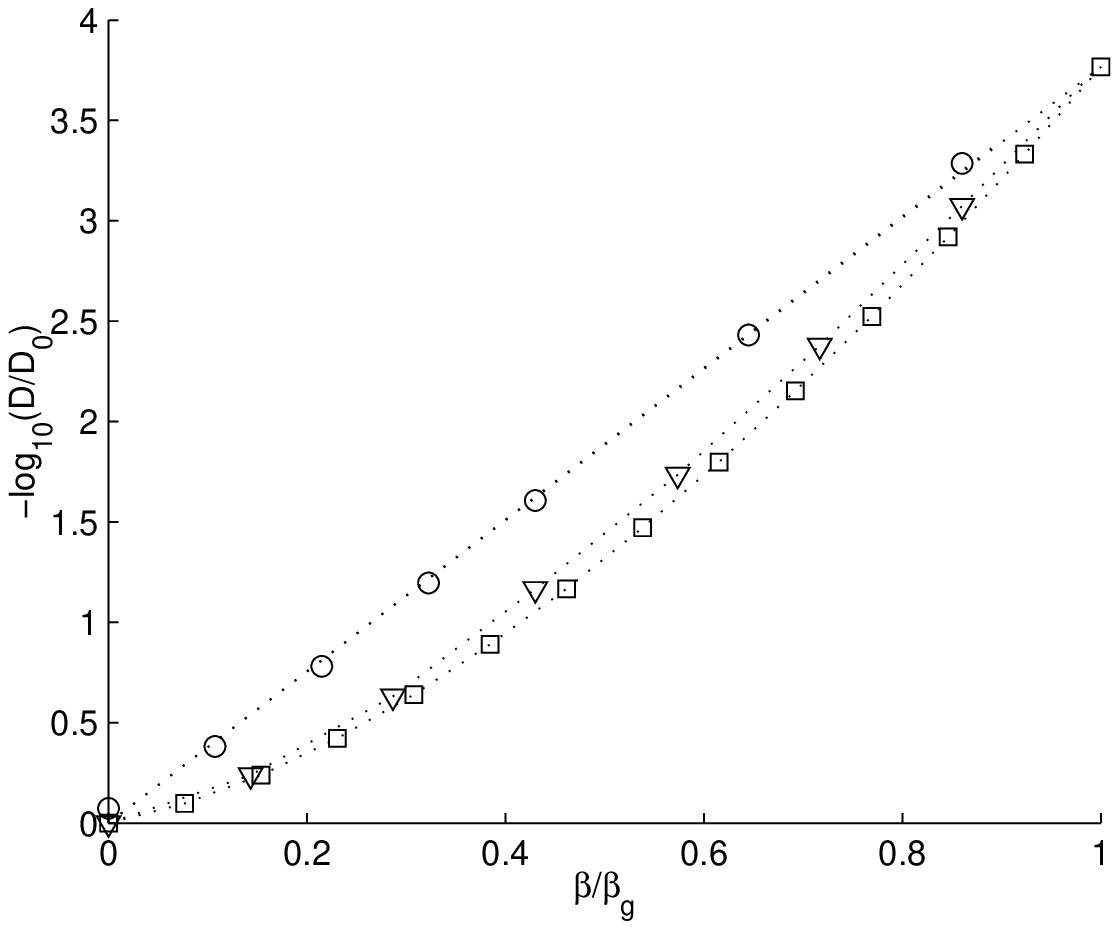}}
\caption{Diffusion coefficients extracted from the mean square
  displacements in Fig. \ref{fig:R2}. For reference the 
diffusion coefficient in the $\rho = 0$ limit (non-interacting
particles) is included. Lower panel: Same data as upper panel. Y-axis
scaled by $D_0\equiv D(\beta = 0)$. X-axis scaled by  
$\beta_g$ to make the data collapse at $\beta/\beta_g = 1$. 
At the density $\rho_h = 1\times 10^{-3}$ we define $\beta_g = 13$.
For $\rho_h = 8\times 10^{-3}$ and $\rho_h = 1$ empirical scaling 
was used to find $\beta_g = 13.95$ and $\beta_g = 30.9$ respectively.
For $\rho = 0$ a straight line was fitted to the data. For the high densities
data points are connected with straight lines.
}
\label{fig:DiffCoeff}
\end{figure}

In Fig. \ref{fig:DiffCoeff}a we report the diffusion coefficients
extracted from Fig. \ref{fig:R2}. For reference we show here also 
results for the $\rho=0$ limit, i.e., simulations with non-interacting 
particles (in this limit the model is obviously not a good model of
a liquid).
In the $\rho=0$ limit we find  Arrhenius
behavior [$D = D_0\exp(-\beta\Delta E)$], as expected from  percolation 
arguments \cite{Perc}. In contrast, the higher densities show distinctive 
non-Arrhenius behavior;  the model exhibits "fragile" behavior.
To facilitate comparison with  Angell's fragility plot \cite{Angell}, we
show in Fig. \ref{fig:DiffCoeff}b the diffusion coefficients scaled
in the following way: the y-axis is scaled with 
the diffusion coefficient at infinite temperature, 
$D_0 \equiv D(\beta = 0)$ (which scales with
$\rho_h$ \cite{vanBeijeren}), and the x-axis is scaled with an 
"inverse glass temperature", $\beta_g$, which is here defined
by the scaled diffusion coefficients being identical at $\beta/\beta_g = 1$ 
(and $\beta_g\equiv 13$ for $\rho_h = 1\times 10^{-3}$).  
The degree of fragility is observed to increase 
with increasing density (decreasing $\rho_h$). 

\begin{figure}
\resizebox{8.5cm}{!}{\includegraphics{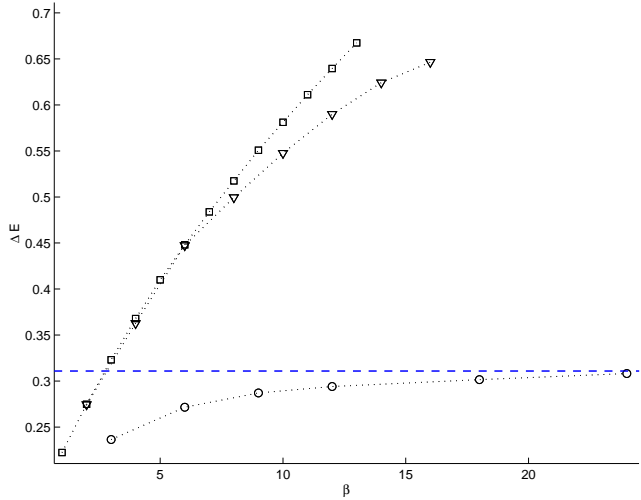}}
\caption{Apparent activation energies calculated from 
eq. \ref{eq:Ea}. 
Squares: $\rho_h = 1 \times 10^{-3}$. 
Triangles: $\rho_h = 8 \times 10^{-3}$.
Circles: $\rho_h \rightarrow 1$ (i.e., $\rho=0$ limit, 
non-interacting particles). 
 Data points are connected with straight lines.
}
\label{fig:Ea}
\end{figure}

Fig. \ref{fig:Ea} shows the apparent activation energy, 
obtained by regarding $\Delta E$ in the Arrhenius expression 
as being temperature dependent:
\begin{eqnarray}
  \label{eq:Ea}
  \Delta E = -\beta^{-1}\log \left( \frac {D}{D_0} \right)
\end{eqnarray}
In the $\rho=0$ limit $\Delta E$ as expected approaches 
the theoretical value, $Ec = 0.31$ \cite{Percolation}. At  high densities, 
$\Delta E$ keeps increasing above this value, reflecting the 
non-Arrhenius behavior. There is an indication (particularly for
$\rho_h = 8\times10^{-3}$) that $\Delta E$ 
might be leveling of to a constant, indicating that there might be 
a crossover from  non-Arrhenius to Arrhenius behavior, as seen 
e.g. in simulations of viscous silica \cite{Horbach99}. 
The observed indication of a crossover to Arrhenius  
behavior might be related to "hitting the bottom" of the energy landscape, 
but this point deserves further investigations. 

\subsection{Density-density correlation (self part)} 

\begin{figure}
\resizebox{8.5cm}{!}{\includegraphics{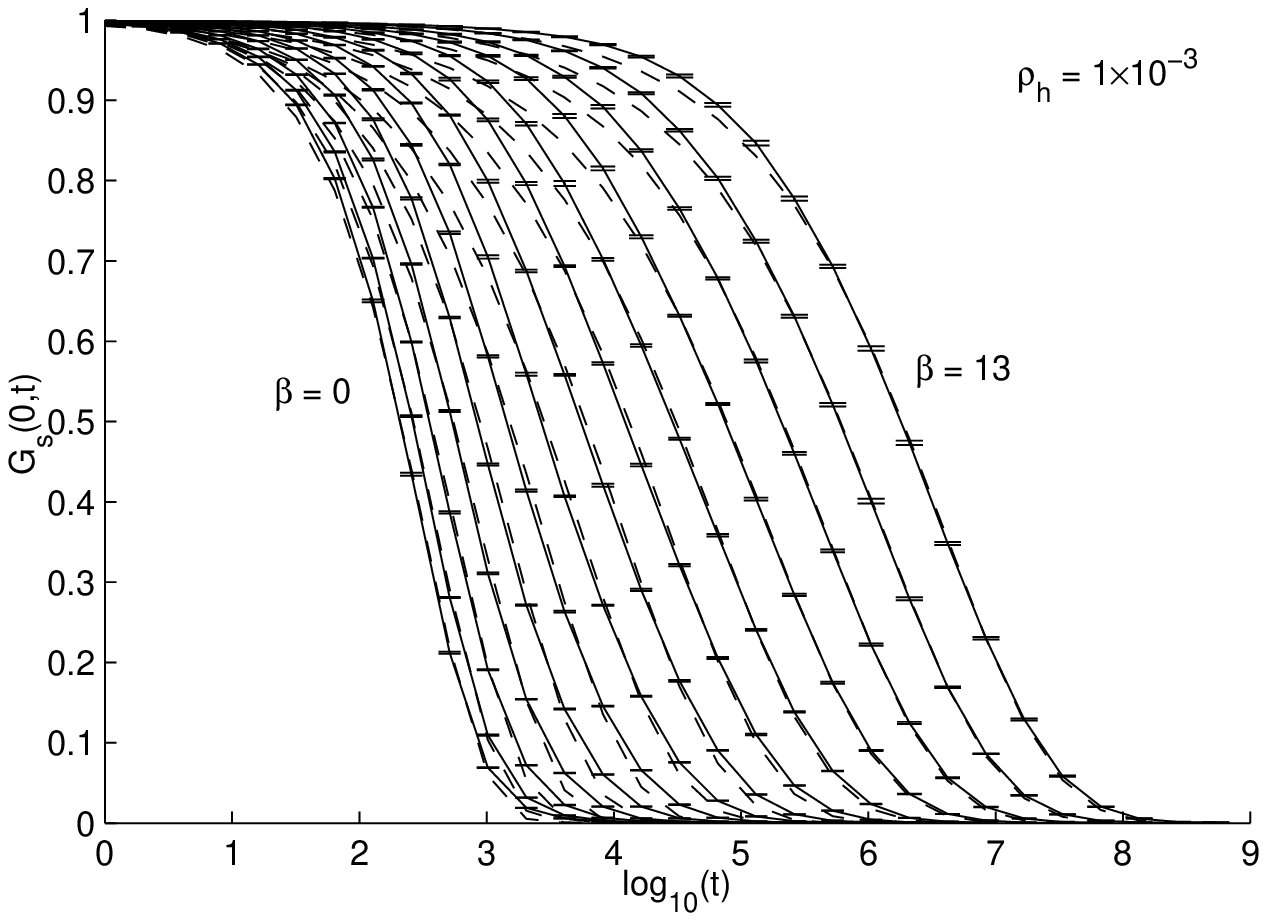}}
\resizebox{8.5cm}{!}{\includegraphics{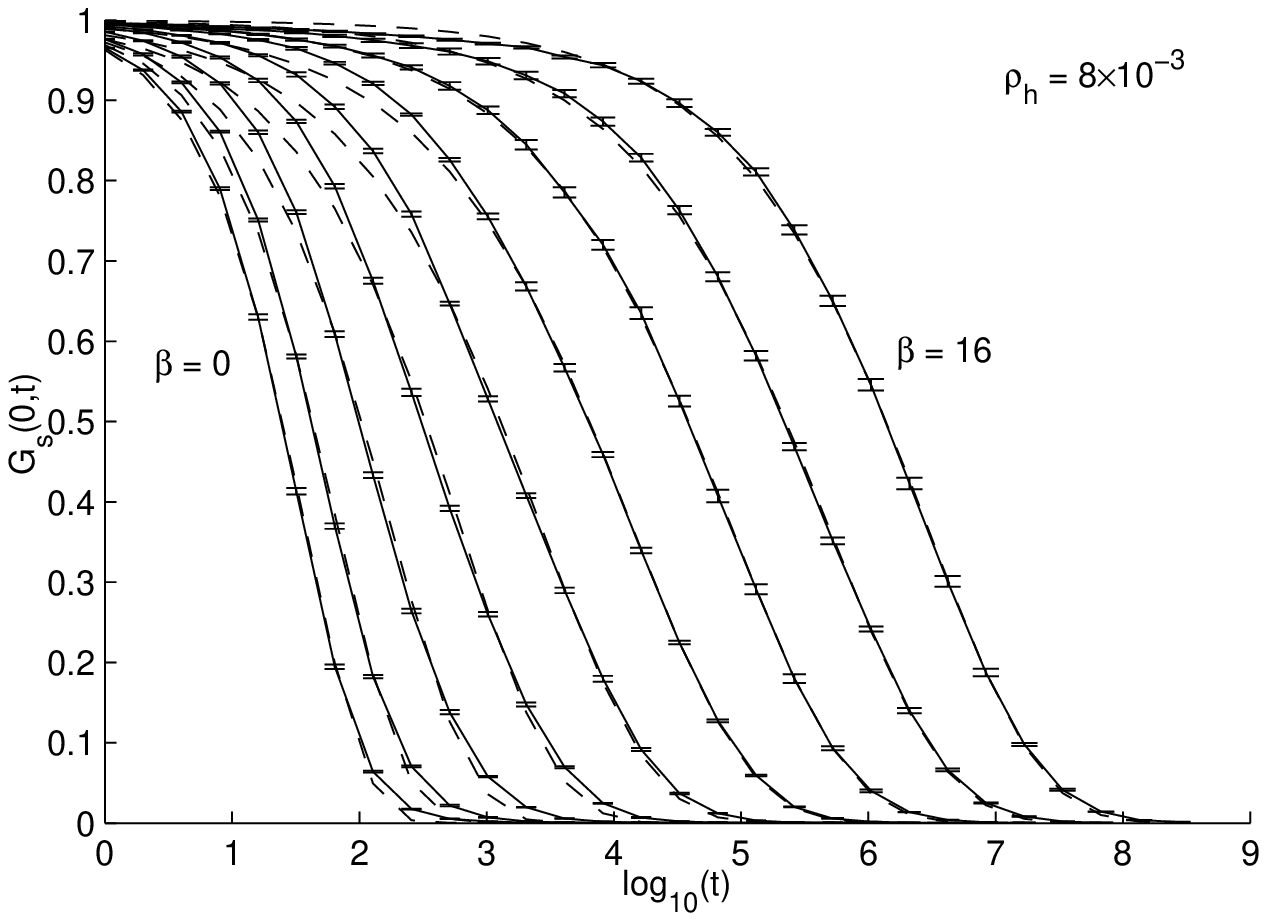}}
\resizebox{8.5cm}{!}{\includegraphics{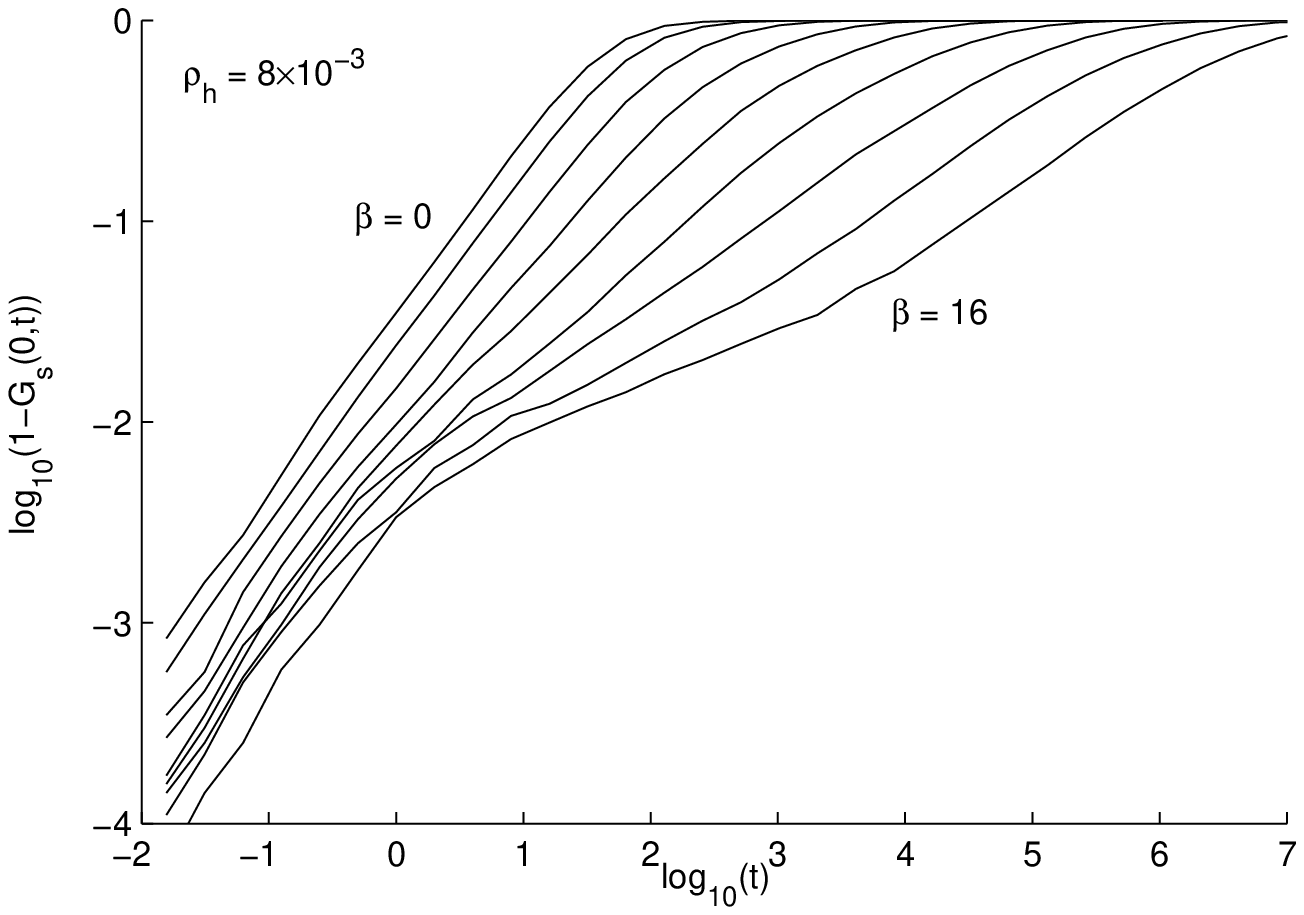}}
\caption{Upper and middle panel: Self part of the density-density
correlation, $G_s(0,t)$ for $\rho_h = 1\times 10^{-3}$ and $\rho_h = 8\times 10^{-3}$ respectively. 
Full lines are straight lines connecting data points. Error-bars indicate 
$95\%$ confidence interval in $G_s(0,t)$ estimated from 
fluctuations between 8 independent samples. 
Dashed lines are fits to stretched exponentials, 
$\exp(-(t/\tau)^\gamma)$. Fits were
done for $G_s(0,t)<0.8$. Lower panel: $1 - G_s(0,t)$ for $\rho_h = 8\times 10^{-3}$.
}
\label{fig:P0}
\end{figure}

\begin{figure}
\resizebox{8.5cm}{!}{\includegraphics{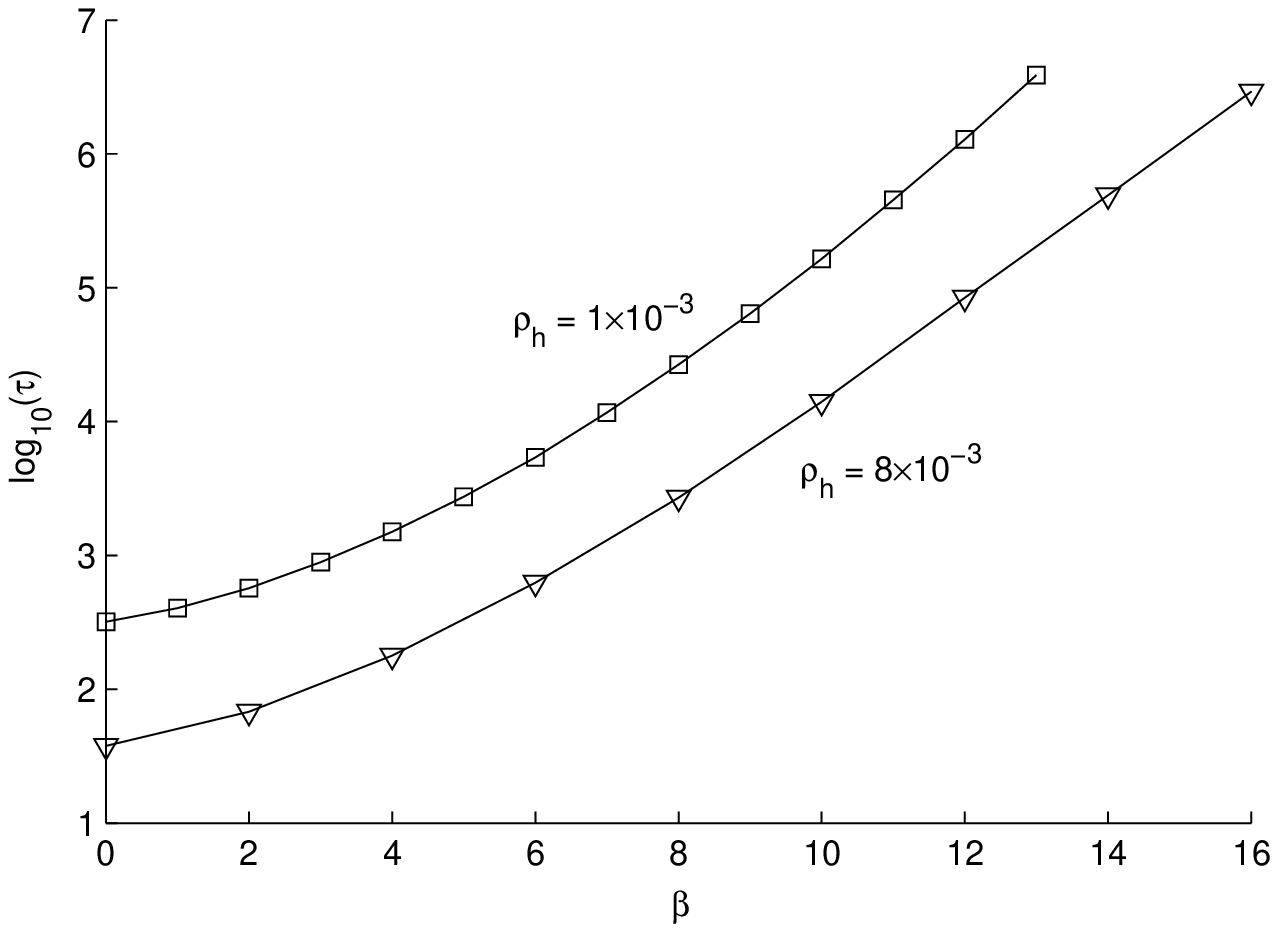}}
\resizebox{8.5cm}{!}{\includegraphics{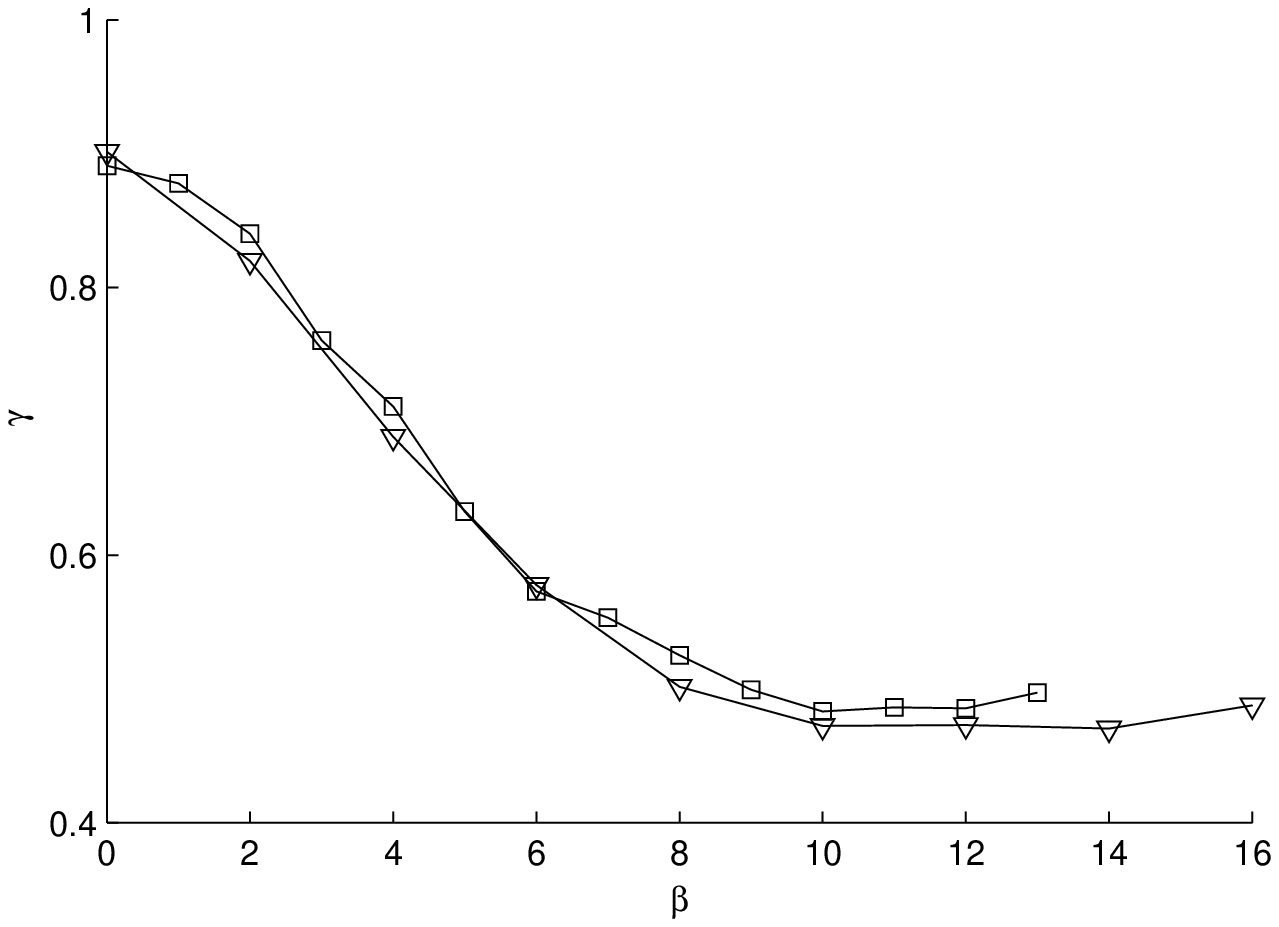}}
\caption{Fitting parameters from fitting stretched exponentials 
to the return probability, $G_s(0,t)$ (fig. \ref{fig:P0}). Upper panel: 
relaxation time $\tau$. Lower panel: stretching exponent
$\gamma$. Data points are connected by straight lines.
}
\label{fig:P0fitparam}
\end{figure}

In Fig. \ref{fig:P0} we show the self part of the density 
auto-correlation, $G_s(0,t)$ \cite{vanHove}, 
i.e., the probability that a particle at time $t$ 
is at the same site as it was at time $t=0$. 
Dashed lines
are fits to stretched exponentials:
\begin{eqnarray}
  \label{eq:StretchedExp}
   f(t) = \exp(-(t/\tau)^\gamma)
\end{eqnarray}
The fits are not perfect, but they  capture the main characteristics 
of the data. The fitting parameters are shown in Fig. 
\ref{fig:P0fitparam}. As for the diffusion constant
(Fig. \ref{fig:DiffCoeff}), the relaxation time $\tau$
exhibits non-Arrhenius behavior with an indication of a crossover to 
Arrhenius behavior at the lowest temperatures. The stretching exponent
$\gamma$ decreases with $\beta$, indicating an increasing degree
of non-exponential relaxation. Except for the lowest temperature at each 
density, the stretching exponent $\gamma$ seems to level off to a
constant close to $0.5$. 
A constant stretching exponent indicates  time-temperature 
superposition (TTS), i.e., that the shape of the relaxation function
is independent of temperature. We note that this behavior is 
consistent with experiments indicating that TTS is correlated to 
$\gamma = 0.5$ \cite{TTS2001}.
Here we find at the very lowest
temperatures an indication that the  stretching exponent starts to 
increase again, which might be related to the apparent cross-over 
from Arrhenius to non-Arrhenius behavior discussed
earlier. Simulations at lower temperatures are needed to investigate
this question further.

Comparing Fig. \ref{fig:R2} to  Fig. \ref{fig:P0}a and  Fig. \ref{fig:P0}b, 
one might ask: ``why does a plateau develop in $\left<r^2(t)\right>$ and not 
in $G_s(0,t)$?''. The answer is, that there is indeed a plateau developing in 
 $G_s(0,t)$ - this can be seen in Fig. \ref{fig:P0}c, where we have plotted
$1-G_s(0,t)$ (i.e. the fraction of particles that are contributing to 
$\left<r^2(t)\right>$) in a log-log plot. In fact, at short times where no particle
jumps more than once $\left<r^2(t)\right> = 1-G_s(0,t)$. At the 
lowest temperature in Fig. \ref{fig:P0}c this relation holds to within $10\%$ for 
$t \leq 100$. 

We note that in the $\rho = 0$ limit at low temperatures $G_s(0,t)$ looks
quantitatively different from what is seen in Fig. \ref{fig:P0}: There is a 
strong initial relaxation (related to jumps with $\Delta E < 0$), a stronger 
stretching ($\gamma \approx 0.3$), and a final pronounced power law regime that starts  at 
$G_s(0,t) \approx 0.1$. This limit (which we again stress is not a good model for 
a liquid) is investigated in a separate paper \cite{SchroderRho0}.

\subsection{Dynamical heterogeneity}


\begin{figure}
\resizebox{8.5cm}{!}{\includegraphics{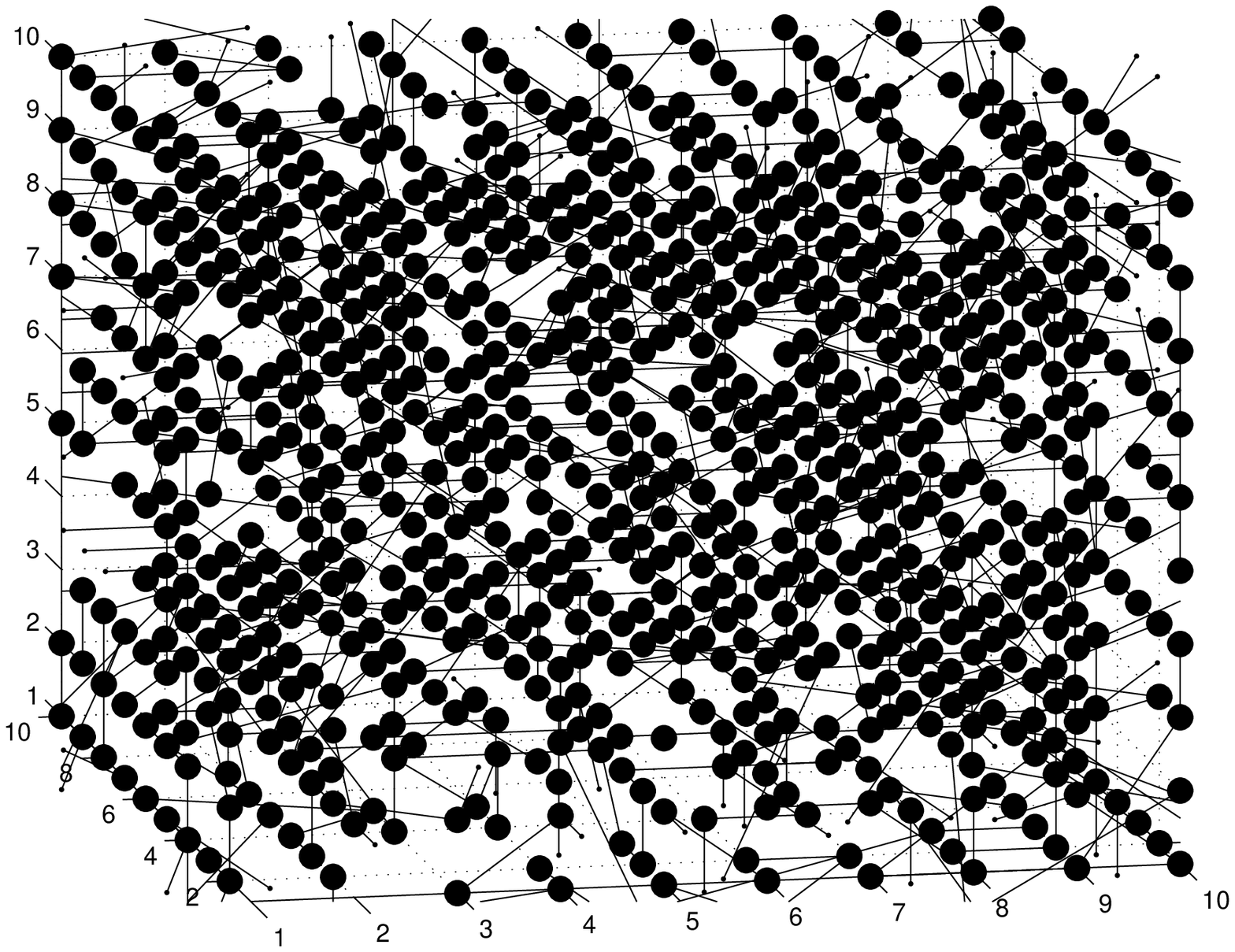}}
\resizebox{8.5cm}{!}{\includegraphics{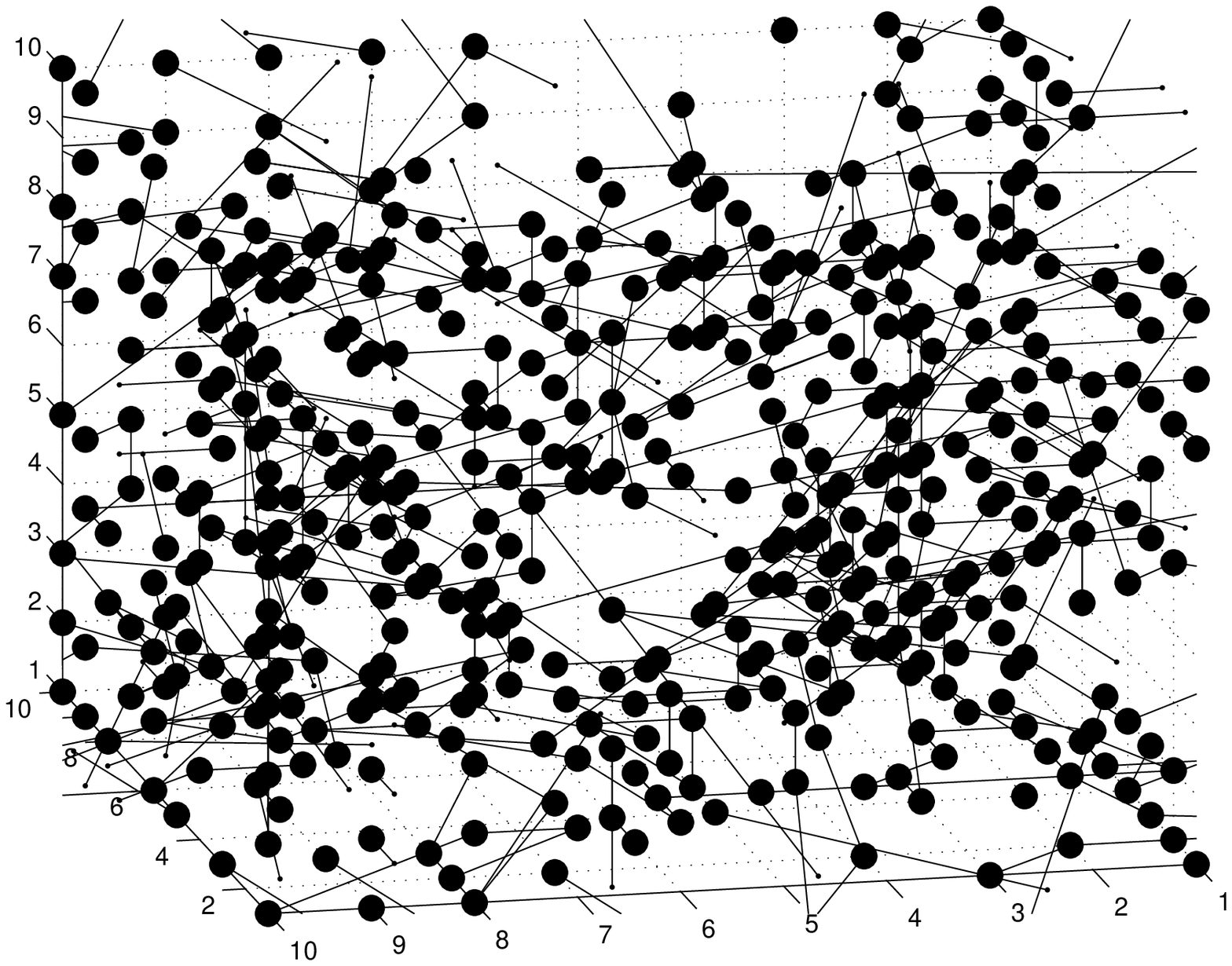}}
\caption{Displacement vectors, at a time where 
$\left< r^2\right> \approx 2$, for $\beta = 0$ and $14$ 
respectively ($\rho = 0.992$, see table \ref{tab:DynHet} for details).
The initial position of particles that moved to a new lattice site 
during the time interval is indicated by a filled circle, and the 
displacement vectors are shown as straight lines.
}
\label{fig:DynHet}
\end{figure}

\begin{table}
\center{
\begin{tabular}{|c|c|c|c|c|} \hline
 $\beta$ & $\left<r^2(t)\right>$ & $t$ & $G_s(0,t)$ \\ \hline
  0  & $2.02$ & $6.2\times 10^1$ & $18.6\%$  \\ \hline
 14  & $2.06$ & $2.0\times 10^5$ & $54.1\%$ \\ \hline
 \end{tabular}
}
\caption{Parameters describing the two different sets of displacement
  vectors in Fig. \ref{fig:DynHet}. Note: Averages are here only
  over particles, \emph{not} ensemble/time averages. 
}
\label{tab:DynHet}
\end{table}

It is well established that viscous liquids contain
dynamical heterogeneities, i.e., if subsets of particles 
are defined by their dynamical properties, these tend to be correlated in 
time and/or space \cite{Sillescu,Ediger}. 
Figure \ref{fig:DynHet} indicates in a qualitative 
way that the model exhibits dynamical heterogeneity to an increasing 
degree as temperature is lowered. 
In Fig. \ref{fig:DynHet} we show  the displacement of particles at a time where
$\left< r^2(t)\right> \approx 2$ for $\beta=0$ and $\beta=14$ 
($\rho_h = 8\times 10^{-3}$). It is evident  from the figure 
that the fraction of particles contributing to the mean
square displacement (i.e. $1 - G_s(0,t)$) is smaller at the low temperature
(see also table \ref{tab:DynHet}), and that the positions of
contributing particles are correlated in space.

\section{Conclusion}

A novel lattice-gas model of viscous liquids with 
extensive average energy and intensive relaxation times and diffusion 
coefficients has been proposed. The first results from simulations of 
the model have been presented. At high densities 
the model exhibits 
the two non's characterizing viscous liquids, 
non-exponential relaxation and non-Arrhenius temperature dependence of
relaxation times and diffusion coefficients. The fragility increases 
with density. Furthermore, indications of a  number of interesting features
was found;
i)  Non-Arrhenius to Arrhenius transition. 
ii) Time-temperature superposition. 
iii) Dynamical heterogeneities.


\acknowledgements{This work was supported by a grant from the Danish National Research      
Foundation for funding the DNRF centre for viscous liquid dynamics "Glass and
Time".}

\printfigures

\printtables


\begin{references}
\bibitem{Perc}
At low temperatures the dynamics are dominated by 
the largest energy a particle has to cross to go through the
sample \cite{dyr95}.
\bibitem{vanHove}
$G_s({\bf r},t)$ is the self part of the van Hove correlation 
function, i.e., the probability that particles are displaced 
the vector ${\bf r}$ in the time interval $t$.
\end{references}
\end{document}